\documentclass[fleqn,twoside]{article}
\usepackage[headings]{espcrc2}

\readRCS
$Id: espcrc2.tex,v 1.2 2004/02/24 11:22:11 spepping Exp $
\usepackage{graphicx}
\usepackage[figuresright]{rotating}

\newcommand{\AmS}{{\protect\the\textfont2
  A\kern-.1667em\lower.5ex\hbox{M}\kern-.125emS}}

\title{Associated W and Higgs boson photoproduction and other electroweak photon induced processes at the \textsc{lhc}}

\author{S. Ovyn\address[MCSD]{Universit\'e catholique de Louvain, Center for Particle Physics and Phenomenology (CP3), Louvain-la-Neuve, Belgium}%
        \thanks{E-mail: severine.ovyn@uclouvain.be.}}
       
\runtitle{Proceedings of workshop on "High energy photon collisions at the LHC"}
\runauthor{S. Ovyn}

\begin{document}

\begin{abstract}
Experimental prospects for studying at the \textsc{lhc} photon-proton interactions at center of mass energies up to and above 1 TeV are discussed. Cross sections are presented for many electroweak and beyond the Standard Model processes. Selection strategies based on photon interaction tagging techniques are discussed. Assuming a typical \textsc{lhc} multipurpose detector, the production of single top associated to a W, and anomalous top signals and their irreducible backgrounds are presented after applying detector acceptance cuts. The associated photoproduction of Higgs and W bosons has a typical cross section of 20~fb. The possibility of observing this reaction is investigated for topologies with signal-to-noise ratio close to unity.
\vspace{1pc}
\end{abstract}

\maketitle

\section{Introduction}
\label{intro}

A significant fraction of $pp$ collisions at the \textsc{lhc} will involve (quasi-real) photon interactions occurring at energies well beyond the electroweak energy scale~\cite{bib:piotr}. The \textsc{lhc} can therefore be considered to some extend as a high-energy photon-proton collider. In a recent paper~\cite{bib:nous}, the initial comprehensive studies of high energy photon interactions at the LHC were reported. In the present contribution, the selected results obtained in~\cite{bib:nous} are introduced and supplemented by new results. Photon interactions can be studied thanks to the experimental signature of events involving photon exchanges: the presence of very forward scattered protons.

Using the equivalent photon approximation (\textsc{epa})~\cite{bib:epa}, the $pp$ cross sections of $pp(\gamma q/g \rightarrow X)Yp$ processes are obtained using the relative luminosity spectra $f_\gamma$:

\begin{equation}\label{equ:a}
\sigma_{pp}=\int \sigma_{\gamma q/g}~f_{\gamma}(x_1)~f_{q/g}(x_2)~dx_1~dx_2 ,
\end{equation}

where $\sigma_{\gamma q/g}$ is the photon-parton cross section, $f_{q/g}$ is the luminosity spectra of the parton.

\begin{figure}[!h]
 \includegraphics[height=.5\textheight]{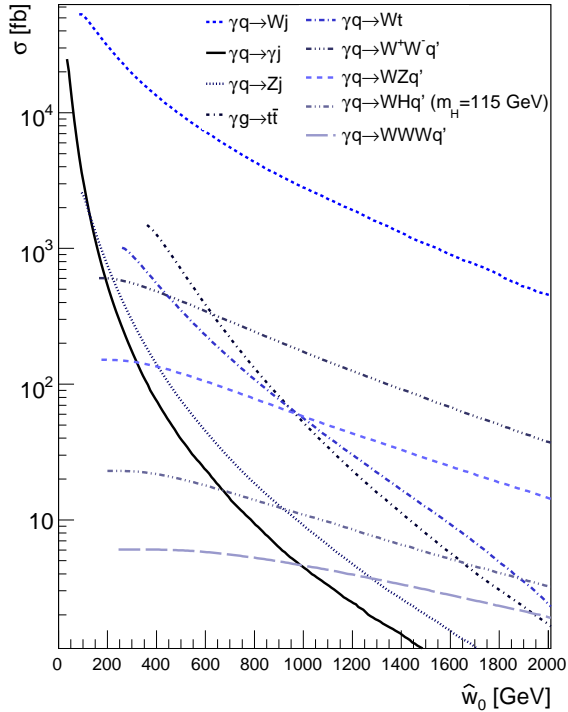}
\caption{Direct contribution at \textsc{lo} for $pp (\gamma q/g \rightarrow X) p Y$ processes as a function of the minimal photon-parton c.m.s. energy $\hat{W}_0$~\cite{bib:nous}. Cross sections have been evaluated using \textsc{mg/me} or \textsc{c}alc\textsc{hep}. For all jets, $p_T^{jet}>$~10~GeV, $|\eta^{jet}|<$~5 and $\Delta R(j,j)>$~0.3.
No other cut than the regularisation cut $p_T >1~$GeV is applied on $q'$}
\label{fig:ovyn_fig1}
\end{figure}

The luminosity and c.m.s. energy of photon-proton collisions are higher than the $\gamma \gamma$ ones~\cite{bib:tomek}. This offers interesting possibilities for the study of electroweak interactions and for searches beyond the Standard Model (\textsc{bsm}) up to TeV scale. Figure~\ref{fig:ovyn_fig1} shows direct photoproduction contribution at \textsc{lo}, evaluated using \textsc{mg/me}~\cite{bib:mad1,bib:mad2} or \textsc{c}alc\textsc{hep}~\cite{bib:calchep}, as a function of the minimal photon-parton c.m.s. energy $\hat{W}_0$. A large variety of $pp(\gamma g/q \rightarrow X)pY$ processes has sizable cross section and could therefore be studied during the very low and low luminosity phases of \textsc{lhc}. Interestingly, potentially dangerous Standard Model background processes with hard leptons, missing energy and jets coming from the production of gauge bosons, have cross sections only one or two orders of magnitude higher than those involving top quarks.

\section{Fast detector simulation}

Photon-proton processes discussed in this paper involve topologies with hard jets in the final state. In order to take into account the effect of jet algorithms and the efficiency of event selection under realistic experimental conditions, the generated events were passed: (1) to \textsc{pythia} 6.227~\cite{bib:pythia} and (2) a fast simulation of a typical \textsc{lhc} multipurpose detector. This simulation assumes geometrical acceptance of sub-detectors and their finite energy resolutions. Electrons and muons are reconstructed if they fall into the acceptance of the tracker ($|\eta|<2.5$) and have a $p_T^{\ell}>$~10~GeV. Jets are reconstructed using a cone algorithm with $R=0.7$ and using the smeared particle momenta. The reconstructed jets are required to have a transverse momentum above 20~GeV and $|\eta^j|<3.0$. A jet is tagged as b-jet if its direction lies in the acceptance of the tracker and if it is associated to a parent b-quark. A b-tagging efficiency of 40$\%$ is assumed. For c-jets and light quark/gluon jets, fake b-tagging efficiencies of 10$\%$ and 1$\%$ respectively are assumed. $\tau$-jets are reconstructed with a typical efficiency of 60$\%$ only if their $p_T^{\tau-jet}$ is higher than 10~GeV.

\section{Tagging and forward proton detectors}
\label{sec.rp}

Tagging is essential for the extraction of high energy photon-induced interactions from the huge $pp$ events. Photon-induced interactions are characterised by a large pseudorapidity region completely devoid of any hadronic activity. This region is usually called {\it large rapidity gap} (\textsc{lrg}). 

\begin{figure}[!h]
 \includegraphics[scale=0.39]{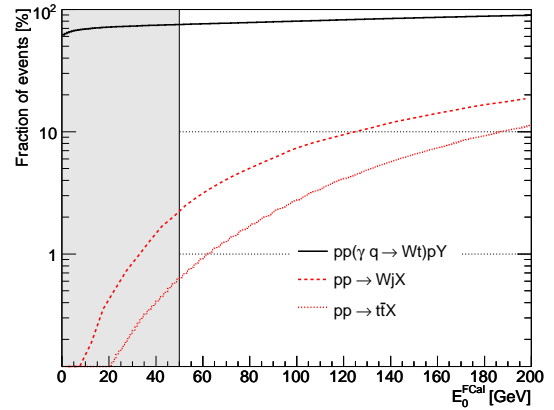}
\caption{\small{Fraction of selected events as a function of the rapidity gap cut 
$E^{FCal}_{0}$ displayed for photon-parton induced $Wt$ and parton-parton induced 
$t\overline{t}$ and $Wj$ final states. $E^{FCal}_{0}$ is defined as the cut on the 
minimal of energies $E^{FCal}_{min}$ measured in the two forward calorimeters. FCal are assumed to cover $3<|\eta|<5$. No other acceptance cut is applied. }}
\label{fig:ovyn_fig2}
\end{figure}

During the phase of low luminosity (i.e. significantly lower than $10^{33}$ cm$^{-2}$s$^{-1}$), the \textit{event pile-up} is negligible. Thanks to the colour flow in $pp$ interactions between the proton remnant and the hard hadronic final states, a simple way to suppress generic $pp$ interactions is to require \textsc{lrg}s. The \textsc{lrg} condition can be applied using a cut based on the energy measured in the forward detector containing the minimum forward activity ($3<|\eta|<5$). For a maximal allowed energy of 50~GeV, a typical reduction factor of 10$^{-3}$ and 10$^{-2}$ for a parton-parton $t\overline{t}$ and $Wj$ production respectively (Figure~\ref{fig:ovyn_fig2}) is expected. This tagging technique, denoted as $E^{FCal}$, is applied to all processes presented in the following sections with an upper cut at 50~GeV. The advantage is that this cut can be done using the central detector only. However, as the energy of the escaping protons is not measured, the event kinematics is less constrained. A total integrated \textsc{lhc} luminosity of 1~fb$^{-1}$ without {\it pile-up} seems to be a realistic assumption.

In this paper, the only considered backgrounds come from photoproduction. However, potentially dangerous backgrounds arise when topologies similar to signal events are produced from the small fraction of parton-parton collisions containing rapidity gaps. The reduction factor due to \textsc{lrg}s might not be sufficient for several $pp$ processes (e.g. Wj) given their very large cross section. The rejection can be further improved by tightening the cut which defines the presence of a rapidity gap (e.g. 30~GeV instead of 50~GeV) and also by using other exclusivity conditions related for instance to the number of tracks. An exclusivity condition requiring no tracks, excluding isolated leptons and jet cones, with $p_T>$~0.5~GeV and 1~$<\eta<$~2.5 in the hemisphere where the rapidity gap is present is applied. With these newly defined acceptance cuts, rapidity gap and exclusivity conditions, efficiency for signal processes drops roughly by a factor of two while the reduction factors for parton-parton reactions are better than $10^{-3}$ (see Table \ref{tab:ovyn_tab1}).

\begin{table}[!h]
\newcommand{\m}{\hphantom{$-$}}
\newcommand{\cc}[1]{\multicolumn{1}{c}{#1}}
\renewcommand{\arraystretch}{1.2} 
\renewcommand{\tabcolsep}{0.8pc} 
\caption{Cross-sections in fb before and after acceptance cuts corresponding to $\ell bjj$ topology. $\mathbf{\sigma}_{visible}$ is the cross section after $E^{FCal}$, acceptance and exclusivity cuts.}
\begin{tabular}[!h]{lccc}
\hline
Topology & $\mathbf{\sigma}$ &  $\mathbf{\sigma}_{FCal}$ & $\mathbf{\sigma}_{visible}$\\
\hline
$pp \rightarrow t\bar{{t}}$ & 328~$\times10^3$ & 791 & 9.96\\
$pp \rightarrow tj$ & 66.6$\times10^3$ & 328 &  0.67\\
\hline
\end{tabular}
\\[1pt]
\label{tab:ovyn_tab1}
\end{table}

Providing good control of the energy scale of forward calorimeters and efficient tagging based on \textsc{lrg}s, one expects inclusive parton-parton processes to be negligible at low luminosity or, at most, of the same order of magnitude than the irreducible photon-induced backgrounds.

At high luminosity, the \textsc{lrg} technique cannot be used because of large event pile-up. Therefore the use of dedicated \textit{very forward detectors} (\textsc{vfd}s)~\cite{bib:xavier} is mandatory in order to retain $pp$ backgrounds low. An integrated luminosity of 100~fb$^{-1}$ is usually assumed in this case.

\section{Associated WH production}

The possibility of using $\gamma p$ collisions to search for $WH$ associated production was already considered at electron-proton colliders~\cite{bib:wh}. At the \textsc{lhc} the cross section for $pp(\gamma q \rightarrow WHq')pY$ reaction reaches 23~fb for a Higgs boson mass of 115~GeV and diminishes slowly down to 17.5~fb with increasing Higgs boson masses up to 170~GeV. Although the cross sections for $\gamma p$ interactions are smaller than the parton-parton ones, the ratio of signal to background cross sections is better in $\gamma p$ processes by more than one order of magnitude (Figure~\ref{fig:ovyn_fig1}).

\begin{figure}[!h]
 \includegraphics[scale=0.38]{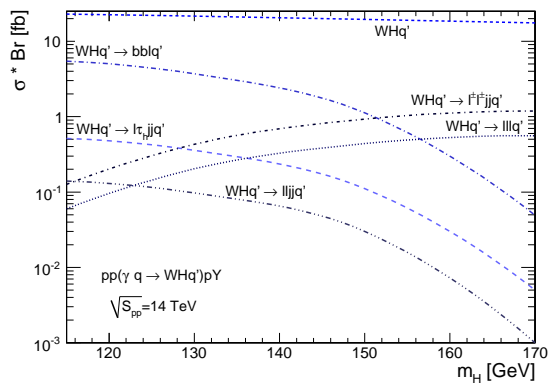}
\caption{The $\gamma p \rightarrow WHq'$ production cross section as well as the cross section times branching ratio for five final states at the \textsc{lhc}.}
\label{fig:ovyn_fig3}
\end{figure}

Five different topologies have been considered for the signal:

\begin{tabular}[!h]{l}
- $W H \rightarrow \ell \nu b\overline{b}$, $\ell = e, \mu, \tau$\\
- $W H \rightarrow W \tau^+ \tau^- \rightarrow jj \ell^+ \ell ^-$, $\ell = e, \mu$\\
- $W H \rightarrow W \tau^+ \tau^- \rightarrow jj \ell \tau_h$, $\ell = e, \mu$\\
- $W H \rightarrow W W^+ W^- \rightarrow \ell \ell \ell$, $\ell = e, \mu, \tau$\\
- $W H \rightarrow W W^+ W^- \rightarrow jj \ell^{\pm} \ell^{\pm}$, $\ell = e, \mu, \tau$. \\
\end{tabular}

\begin{table}[!h]
\begin{center}
\newcommand{\m}{\hphantom{$-$}}
\newcommand{\cc}[1]{\multicolumn{1}{c}{#1}}
\renewcommand{\arraystretch}{1.2} 
\renewcommand{\tabcolsep}{0.4pc} 
\caption{\small{Acceptance cuts for the five topologies resulting from $pp(\gamma q \rightarrow WHq')pY$ process.}}
\label{tab:ovyn_tab2}
\begin{tabular}[!h]{c c c c c c}
\hline & $\ell b\overline{b}$ & $jj \ell \ell$ & $jj \ell \tau_h $ & $\ell \ell \ell$ & $\ell^{\pm} \ell^{\pm} jj$\\\hline
$\mathrm{N_{\ell}}$ & 1 & 2 & 1 & 3 & 2\\
$\mathrm{N_{\tau_h}}$ & - & - & 1 & - & -\\
$\mathrm{N_{jet}}$ & 2 $b$-tag&  2 & 2 &$\leq 1$ & $\geq 2$ \\
$\mathrm{|\eta^{jet}_{max}|}$ & 3 & 3 & 3 & 3 & 3\\
\hline
\end{tabular}
\end{center}
\end{table}

The $WHq'$ production cross section as well as the cross section times branching ratio for the five topologies are summarised in Figure~\ref{fig:ovyn_fig3}. We analyze the topologies arisen from $H\rightarrow b\overline{b}$ and $H\rightarrow\tau^+\tau^-$ for a Higgs mass of 115~GeV. The two final states obtained from the $H\rightarrow W^+W^-$ decay are studied for a heavier Higgs boson: $M_H$~=170~GeV. The $\gamma p$ events considered as irreducible backgrounds are: $t\overline{t}$, $Wt$, $Wb\overline{b}q'$, $W\ell \ell q'$, $WZq'$ and $WWWq'$. For each topology, acceptance cuts arisen directly from the final state (see Table~\ref{tab:ovyn_tab2} are applied. The visible cross sections after their applications, summarised in Table~\ref{tab:ovyn_tab3} are small. Therefore $WH$ photoproduction could not be considered as a discovery channel. Nevertheless, three channels are promising due to the very good signal to background ratio obtained after the application of very simple acceptance cuts: $\ell b\overline{b}$, $\ell \ell \ell$ and $\ell^{\pm} \ell^{\pm} jj$. Prospects for the observability of these topologies, including  additional photon-induced processes with different final state particles than the signal (called {\it reducible background}) are presented in the three following sections.

\begin{table}[!h]
\caption{Cross sections in fb for five $WHq'$ induced final states before and after acceptance cut 
together with the cross sections of irreducible background processes after acceptance cuts.}
\label{tab:ovyn_tab3}
\newcommand{\m}{\hphantom{$-$}}
\newcommand{\cc}[1]{\multicolumn{1}{c}{#1}}
\renewcommand{\arraystretch}{1.2} 
\renewcommand{\tabcolsep}{1.2pc} 
\begin{tabular}{l c c c}
\hline
  Topology & $\sigma$ & $\sigma_{acc}$ & $\sigma_{acc}^{Bkg}$\\
\hline
$\ell b\overline{b}$ & 5.42 & 0.12 & 3.73\\
$jj \ell \ell$ & 0.14 & 0.01 & 30.8\\
$jj \ell \tau_h $ & 0.52 & 0.04 & 7.56\\
$\ell \ell \ell$ & 0.55 & 0.07 & 1.44\\
$\ell^{\pm} \ell^{\pm} jj$ & 1.17 & 0.22 & 0.28\\
\hline
\end{tabular}
\\[2pt]
\end{table}

\subsection{$W H q' \rightarrow \ell \nu b\overline{b}q'$ topology}

\begin{figure}[!h]
 \includegraphics[scale=0.37]{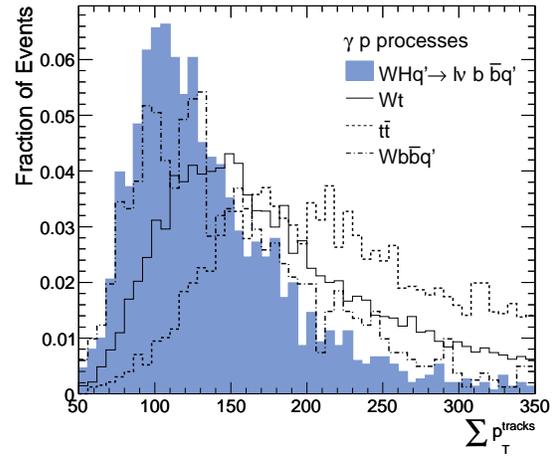}
\caption{Distribution of the scalar sum of the transverse momentum of the isolated lepton and the three jets after the application of the previous analysis cuts for the signal and the photon-induced backgrounds. The distributions are normalised to unity.}
\label{fig:ovyn_fig4}
\end{figure}

The signal final state consists of a lepton from the W decay and a pair of b-jets which gives a peak in the invariant mass distribution. In addition to the presented $E^{FCal}$ cut, events are selected if they contain one isolated lepton with $p_{T}^{\ell}>15$~GeV in the pseudorapidity interval $|\eta^{\ell}|<2.5$ and exactly 2 b-tagged jets with $p_{T}^{b-jet}>$~20~GeV and $|\eta^{b-jet}|<2.5$. The lepton is defined to be isolated if there is no other track with $p_{T}> 2$ GeV in a cone of $\Delta R < 0.5$ around the lepton. Moreover, the events are accepted if they do not contain any additional jet with $p_{T}^j>20$ GeV and $|\eta^j|<3$. The application of this cut is a good method to reduce backgrounds from $t\overline{t}$ and $Wt$ events which usually come with at least one additional jet. An additional topological cut based on the central transverse momentum calculated from the scalar sum of the $p_T$: the obtained value must be comprised between 45 and 140 GeV. This cut has a rejection against $t\overline{t}$ background (see Figure~\ref{fig:ovyn_fig4}).

\begin{table}[!h]
\caption{Cross sections in fb for $W H q'\rightarrow \ell b \overline{b}q'$ final state before and after application of analysis cuts together with the photon-induced background processes after analysis cuts.}
\label{tab:ovyn_tab4}
\newcommand{\m}{\hphantom{$-$}}
\newcommand{\cc}[1]{\multicolumn{1}{c}{#1}}
\renewcommand{\arraystretch}{1.2} 
\renewcommand{\tabcolsep}{1pc} 
\begin{tabular}{l c c c }
\hline
Event & $\sigma$ & $\sigma_{FCal}$ & $\sigma_{Final}$\\
\hline
$WHq'$ &5.42 & 4.77  & 0.06\\
$t\overline{t}$ & 672 & 542  & 0.12\\
$Wt$ & 365 & 268 & 0.12\\
$Wb\bar{b}q'$ & 14.72 & 12& 0.08\\
\hline
\end{tabular}
\\[2pt]
\end{table}

The estimated efficiencies are converted to the final cross sections by multiplying the production cross sections of each processes. The results are summarised in Table~\ref{tab:ovyn_tab4}. In order to exploit the discriminative power from the invariant mass of the two b-tagged jets, the log-likelihood method is used. After 100~fb$^{-1}$ of integrated luminosity, a significance of 1.5~$\sigma$ is obtained. If a discovery is not expected in this channel, a high integrated luminosity could  allow to probe $Hbb$ coupling for a light Higgs boson, which is known to be very challenging to assess in parton-parton processes. 

\subsection{$W H q'\rightarrow W W^+ W^-q'$ topologies}

The events provide a distinctive signature with three $W$ bosons in the final state. It is worth studying the fully leptonic final state in which all the $W$ bosons decay into lepton pairs and the topology in which two of them decay leptonically and one hadronically. In the latest topology, the choice of the like-sign lepton pairs is very useful to reject many background processes with a two-lepton final state signature as $t\overline{t}$ events followed by leptonic decay of both $W$'s from the $t$'s.

\subsubsection{$W W^+ W^- q' \rightarrow \ell \ell \ell q'$ topology}

All Standard-Model processes likely to produce three leptons must be considered as background, including events with a fake lepton. In the present analysis, we considered the photoproduction of $WWWq'$, $W\ell\ell q'$, $Wt$ and $t\overline{t}$ events. Because in the $t\overline{t}$ and $Wt$ events contain a lepton from the semileptonic decay of a B-meson, one of the three leptons is less isolated. Top production is characterised by large jet activity and the three leptons are usually accompanied by at least one b-jet. On the other hand, $WZq'$ background can be strongly rejected by vetoing events which have at least one pair of opposite-sign, same flavor leptons with an invariant mass compatible with the $Z$-mass.

\begin{figure}[!h]
 \includegraphics[scale=0.36]{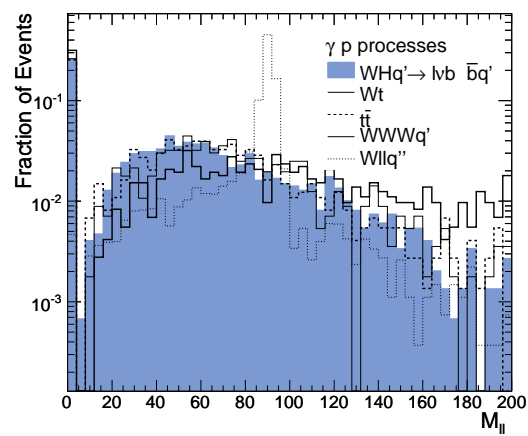}
\caption{Distribution of the invariant mass of the opposite-charge same flavor leptons after the application of the previous analysis cuts for the signal and the photon-induced backgrounds. The distributions are normalised to unity.}
\label{fig:ovyn_fig5}
\end{figure}

Based on these characteristics, the following selection criteria are applied: presence of three isolated leptons (electrons or muons) with $p_T^e>$~15~GeV for electrons and $p_T^{\mu}>$~10~GeV for muons in the pseudorapidity interval $|\eta^{\ell}|<2.5$; events with at least two jets with $p_T^j>$~20~GeV in the region $|\eta^j|< 3 $ are discarded; events containing a b-labeled jet are rejected. Moreover, events which contain at least one pair of opposite-charge same-flavor leptons with an invariant mass between 60 and 120 GeV are rejected (Figure~\ref{fig:ovyn_fig5}). Out of the three possible lepton pairs, at least one should fulfill the $|\Delta \phi| <\pi/2$, $|\Delta \eta|<1$ and $M_{\ell\ell}<$~80~GeV requirements ({\it \textsc{hww}-like} leptons). Finally an event is rejected if the leading jet has $|\eta^j|<1$. The $E^{FCal}$ cut allowing to tag photon-induced events is applied.

\begin{table}[!h]
\caption{Cross sections in fb for $W H q'\rightarrow W W^+ W^- q'\rightarrow \ell \ell \ell q'$ final state before and after application of analysis cuts together with the photon-induced background processes after analysis cuts.}
\label{tab:ovyn_tab5}
\newcommand{\m}{\hphantom{$-$}}
\newcommand{\cc}[1]{\multicolumn{1}{c}{#1}}
\renewcommand{\arraystretch}{1.2} 
\renewcommand{\tabcolsep}{1pc} 
\begin{tabular}{l c c c}
\hline
Event & $\sigma$ & $\sigma_{FCal}$ & $\sigma_{Final}$\\
\hline
$WHq'$            & 0.56 & 0.50 & 0.02\\
$t\overline{t}$   & 159  & 137  & 0.003\\
$Wt$              & 104  & 90   & 0.007\\
$W\ell^+\ell^-q'$ & 12.6 & 11   & 0.016\\
$WWWq'$           & 0.20 & 0.19 & 0.03\\
\hline
\end{tabular}
\\[2pt]
\end{table}

The visible cross sections after the application of this selection procedure are summarised in Table~\ref{tab:ovyn_tab5}. Using the invariant mass of the two {\it \textsc{hww}-like} leptons as discriminant variable, the significance reach 1.6~$\sigma$ after 100~fb$^{-1}$ of integrated luminosity. 

\subsubsection{$W W^+ W^-q' \rightarrow jj \ell^{\pm} \ell^{\pm}q'$ topology}

\begin{figure}[!h]
\includegraphics[scale=0.36]{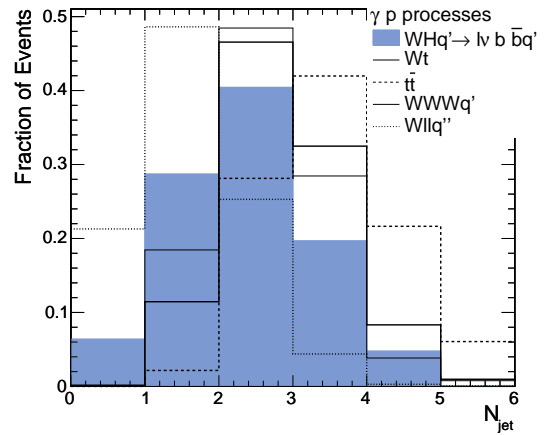}
\caption{Distribution of the number of jets with $p_T^j>$20~GeV and $|\eta^j|<$~3 after the application of the previous analysis cuts for the signal and the photon-induced backgrounds. The distributions are normalised to unity.}
\label{fig:ovyn_fig6}
\end{figure}

When the decay branching ratio of the Higgs boson into $W$ pair becomes dominant, the same sign lepton signature coming from leptonic decays of two out of the three produced $W$ seems to be very promising. After the application of acceptance cuts, this topology has a signal to irreducible background ratio close to one which is unique at \textsc{lhc}. The photo-produced events considered as backgrounds are identical to the one used in the fully leptonic topology. The signal is expected to contain two charged leptons in the final state. Because no other leptons are expected, the events with exactly two same-sign leptons (electrons with $p_T^e>15$~GeV and muons with $p_T^{\mu}>10$~GeV) in the tracker region are conserved. Accepting the  events requires at least two jets with $p_T^{jet}>$~20~GeV and pseudorapidity $|\eta|<$~3. The distribution of the number of these jets is shown in Figure~\ref{fig:ovyn_fig6}. All events containing either a $\tau$-jet, either a b-jet are discarded. These requirements suppress effectively the $WZq'$, $t\overline{t}$ and $Wt$ events. 

\begin{table}[!h]
\caption{Cross sections in fb for $W H q'\rightarrow W W^+ W^- q'\rightarrow jj \ell^{\pm} \ell^{\pm}q'$ final state before and after application of analysis cuts together with the photon-induced background processes after analysis cuts.}
\label{tab:ovyn_tab6}
\newcommand{\m}{\hphantom{$-$}}
\newcommand{\cc}[1]{\multicolumn{1}{c}{#1}}
\renewcommand{\arraystretch}{1.2} 
\renewcommand{\tabcolsep}{1pc} 
\begin{tabular}{l c c c}
\hline
Event & $\sigma$ & $\sigma_{FCal}$ & $\sigma_{Final}$\\
\hline
$WHq'$            & 16.9 & 12   & 0.19\\
$t\overline{t}$   & 672  & 543  & 0.46\\
$Wt$              & 360  & 265  & 0.12\\
$W\ell^+\ell^-q'$ & 8.46 & 7.04 & 0.13 \\
$WWWq'$           & 1.27 & 1.09 & 0.09\\
\hline
\end{tabular}
\\[2pt]
\end{table}

Table~\ref{tab:ovyn_tab6} shows the visible cross section after the application of the analysis cuts. The total number of surviving signal after 100~fb$^{-1}$ is 19. The significance of this topology is 2.6~$\sigma$. Combining the two $H\rightarrow W^+W^-$ topologies, a significance close to 3 can be reached. These channels are important to be studied because they are one of the few signatures for fermiophobic Higgs boson model. In addition, they are doubly dependent on the \textsc{hww} coupling.

\section{Single top Associated to a W}

Photoproduction of single top is dominated by t-channel amplitudes when the top quark is produced in association with a $W$ boson. In contrast to proton-proton deep inelastic scattering where the ratio of $Wt$ associated production cross section to the sum of all top production cross sections is only about $5\%$, it is about 10 times higher in photoproduction. This provides a unique opportunity to study this reaction at the start phase of the \textsc{lhc}. While the overall photoproduction of top quark is sensitive to the top quark electrical charge, the $Wt$ associated photoproduction amplitudes are all proportional to the \textsc{ckm} matrix element $|V_{tb}|$.

The $\gamma p \rightarrow Wt$ process results in a final state of two on-shell $W$ bosons and a $b$ quark. The studied topologies are $\ell bjj$ for the semi-leptonic decay of the two $W$ bosons and $\ell \ell b$ for the di-leptonic decay. The dominant irreducible background of both channels is expected to stem from the $t\overline{t}$ production, where a jet is not identified. Other $\gamma p$ backgrounds are $Wjjj$ and $WWq'$ processes. 

\begin{table}[!h]
\caption{Cross sections in fb for two $Wt$ induced final states before and after acceptance cuts
 together with the cross sections of irreducible background processes after acceptance cuts.}
\label{tab:ovyn_tab7}
\newcommand{\m}{\hphantom{$-$}}
\newcommand{\cc}[1]{\multicolumn{1}{c}{#1}}
\renewcommand{\arraystretch}{1.2} 
\renewcommand{\tabcolsep}{1.2pc} 
\begin{tabular}{l c c c}
\hline
  Topology & $\sigma$ & $\sigma_{acc}$ & $\sigma_{acc}^{Bkg}$\\
\hline
$\ell b jj$   & 440  & 34.1$^{(1)}$ & 63.0\\
$\ell \ell b$ & 104 & 8.69$^{(2)}$ & 3.00\\
\hline
\end{tabular}
\\[2pt]
$^{1}$Maximum $\eta$ for jets is 3\\
$^{2}$Maximum $\eta$ for jets is 2.5 \\
\end{table}

Table~\ref{tab:ovyn_tab7} shows the $pp(\gamma b \rightarrow Wt)Yp$ cross sections before and after the application of acceptance cuts. The visible cross sections of the irreducible backgrounds are also given. The inclusive single top cross section after acceptance cuts of 34~fb, with a signal over irreducible background close to 0.6, suggests an easy discovery of this production mechanism with an integrated luminosity of about 1~fb$^{-1}$. Furthermore, a reduction of the background can easily be obtained by adding more specific analysis cuts. However, a more detailed study would be required to also take into account reducible photoproduction backgrounds and inclusive $pp$ interactions~\cite{bib:jerome}.

\section{Anomalous top production}

\textsc{fcnc} appear in many extensions of the Standard Model, such as two Higgs-doublet models or R-Parity violating
 supersymmetry. The observation of a large number of single top events at the \textsc{lhc} would hence be a clean
 signature of \textsc{fcnc} induced by processes beyond the Standard Model. The effective Lagrangian for this
 anomalous coupling can be written as~\cite{bib:eff_lag_anotop}:

\begin{eqnarray}
\mathcal{L} & = & iee_t\bar{t}\frac{\sigma_{\mu\nu}q^{\nu}}{\Lambda}k_{tu\gamma}uA^{\mu}\\
 & & + iee_t\bar{t}\frac{\sigma_{\mu\nu}q^{\nu}}{\Lambda}k_{tc\gamma}cA^{\mu} + h.c.,
\end{eqnarray}
where $\sigma^{\mu\nu}$ is defined as $(\gamma^{\mu} \gamma^{\nu} - \gamma^{\nu} \gamma^{\mu})/2$, $q^{\nu}$ being the photon 4-vector and $\Lambda$ an arbitrary scale, conventionally taken as the top mass. The couplings k$_{tu\gamma}$ and k$_{tc\gamma}$ are real and positive such that the cross section takes the form:
\begin{equation}
\sigma_{pp \rightarrow t} = \alpha_u\ k^2_{tu\gamma} + \alpha_c\ k^2_{tc\gamma}. 
\end{equation}
The computed $\alpha$ parameters obtained using \textsc{c}alc\textsc{hep} are $\alpha_u = 368$~pb
 and $\alpha_c = 122$~pb. The present upper limit on $k_{tu\gamma}$ is around 0.14, depending on the top mass~\cite{bib:zeus_st} while the anomalous coupling $k_{tc\gamma}$ has not been probed yet.

\begin{table}[!h]
\newcommand{\m}{\hphantom{$-$}}
\newcommand{\cc}[1]{\multicolumn{1}{c}{#1}}
\renewcommand{\arraystretch}{1.2} 
\renewcommand{\tabcolsep}{0.7pc} 
\caption{Cross sections in fb for one anomalous top induced final state ($k_{tu\gamma}$ = 0.1,
 $k_{tc\gamma}$ = 0) before and after acceptance cut together with the cross sections
 of irreducible background processes after acceptance cuts.}
\begin{tabular}[!h]{l c c c }
\hline
Event          & $\ell b$ signal & $Wj$ & $Wc$\\
\hline
$\sigma$  & 769.0 & 53.1$\times10^3$ &  11.4$\times10^3$              \\
$\sigma_{acc}$              & 144.0 &56.2&82.8   \\
\hline
\end{tabular}
\\[2pt]
\label{tab:ovyn_tab8}
\end{table}

The final state is composed of a $b$-jet and a $W$ boson. The studied topology is therefore $\ell b$. Main irreducible background processes come from photoproduced $Wj$ and $Wc$. After the application of acceptance cuts ($p_T^{\ell} >$~10~GeV, $p_T^j >$~20~GeV, $|\eta^{jet,\ell}|<2.5$ and b-tagging of the jet) the cross sections obtained for the signal and for the irreducible backgrounds are similar (Table~\ref{tab:ovyn_tab8}). For the signal, a value of 0.1 was chosen for $k_{tu\gamma}$ while $k_{tc\gamma}$ was set at zero. Due to the large number of events produced and the signal to background ratio close to one, it appears that current limits on the anomalous couplings could be easily improved already after a few months of run of the \textsc{lhc}.

\section{Summary and perspectives}

A survey of several high energy $\gamma$p interactions at \textsc{lhc} has been presented. The high cross section as well as the usually much lower backgrounds offers an ideal framework for studying massive electroweakly interacting particles in a complementary way to the usual, parton-parton processes. Interesting studies and searches can already be performed at the initial integrated luminosity of about one inverse femtobarn. The $Wt$ photoproduction is also surprisingly large, offering an opportunity to measure $|V_{tb}|$ element of the quark mixing matrix. Anomalous $\gamma qt$ couplings might also be uniquely revealed by photoproduction. Larger integrated luminosity, of about hundred inverse femtobarn, will open complementary ways to access important information on the Higgs boson coupling to $b$ quarks and $W$ bosons.



\end{document}